\documentclass[aps,prx,reprint,letterpaper,showpacs,twocolumns]{revtex4-2}
\usepackage{graphicx}
\usepackage{amsmath}
\usepackage{amsfonts}
\usepackage{bm}
\usepackage[colorlinks,urlcolor=blue,anchorcolor=blue,linkcolor=blue,citecolor=blue,breaklinks=true]{hyperref}

\graphicspath{{./newfigures/}{./figures/}}

\begin{document}

\title{Bosonic fractional Chern insulating state at integer fillings in multi-band system}

\author{Wei-Wei Luo$^{1}$}
\author{Ai-Lei He$^{2}$}
\author{Yi-Fei Wang$^{3}$}
\author{Yuan Zhou$^{1,4}$}
\email{zhouyuan@nju.edu.cn}
\author{Chang-De Gong$^{3,1}$}
\affiliation{
    $^{1}$National Laboratory of Solid State Microstructures and Department of Physics, Nanjing University, Nanjing 210093, China \\
    $^{2}$Institute for Advanced Study, Tsinghua University, Beijing 100084, China \\
    $^{3}$Center for Statistical and Theoretical Condensed Matter Physics, and Department of Physics, Zhejiang Normal University, Jinhua 321004, China\\
    $^{4}$Department of Mathematics and Physics, Xinjiang Teacher's College, Urumqi 830043, China
    }
\date{\today}
\begin{abstract}
    The integer quantum Hall state occurs when the Landau levels are fully occupied by the fermions, while the fractional quantum Hall state usually emerges when the Landau level is partially filled by the strongly correlated fermions or bosons. Here, we report two fractional Chern insulating states of the hard-core bosons in a multi-band lattice model hosting topological flat bands with high Chern number. The previously proposed $\nu=1/3$ fractional Chern insulating state inherited from the high Chern number $C=2$ of the lowest topological flat band is revisited by the infinite density matrix renormalization group algorithm. In particular, we numerically identify a bosonic $1/2$-Laughlin-like fractional Chern insulating state at the integer fillings. We show two lower topological flat bands jointly generate an effective $C=1$ Chern band with half-filling. Furthermore, we find a strictly particle-hole-like symmetry between the $\nu$ and $3-\nu$ filling in our model. These findings extend our understanding of quantum Hall states and offer a new route to realize the novel fractional states in the system with multi-bands and high-Chern numbers.
\end{abstract}

\maketitle
{\it Introduction.---}The emergence of topological state and topological order opens a new window to classify phases of matter beyond the Landau's symmetry-breaking paradigm~\cite{wen1990}. The precise quantization of Hall conductivity $\sigma_{xy}=C\frac{e^2}{h}$ with $C$ a topological invariant Chern number appears when the Landau levels generated by the external magnetic field are fully filled. In contrast, the fractional quantum Hall state is observed when the strongly interacting electrons partially fill a Landau level, whose topological order can be characterized by the ground state degeneracy on a compact geometry and the gapless edge excitations described by the chiral Luttinger liquid~\cite{wen1990,wen1995}.
Laughlin proposed a trial wave function that captures the essence of the fractional quantum Hall state, in which the quasiparticle excitations host the fractional charges and fractional braiding statistics~\cite{laughlin1983}. Halperin further generalized the Laughlin state into the two-component system (noted as Halperin ($mmn$) state) to account for the spin or pseudospin degrees of freedom~\cite{halperin1983}. The lattice version of integer quantum Hall state in the absence of the Landau level, also known as the Chern insulating state, was first proposed on a honeycomb lattice model with the staggered flux threading~\cite{haldane1988}, and experimentally realized in the solid state materials~\cite{chang2013} and cold atom systems~\cite{jotzu2014}. The Chern insulators, compared with their continuum counterpart, can survive in a less strict setup and host more exotic topological phases~\cite{ge2020}, and thus provide a promising application platform. Since the $C=1$ Chern band can be adiabatically connected to the Landau level~\cite{liu2013c,scaffidi2012,wu2012c}, the fractional quantum Hall effect was possible to be realized in the lattice models. It was previously proposed for the ultracold atoms confined in optical lattice~\cite{sorensen2005,palmer2006,hafezi2007}, in which the effective magnetic field was simulated by introducing oscillating quadrupole potential or lattice rotation. The proposal of topological flat band (TFB)~\cite{neupert2011,sun2011,tang2011} in the absence of the external magnetic field opens new window to realize the fractional Chern insulator. Subsequently, a series of fractional Chern insulators were established in various lattice models hosting the $C=1$ TFB~\cite{sheng2011,wang2011,regnault2011}.

Unlike the unit Chern number of the Landau level in the continuum limit, the lattice model can host high Chern numbers, such as in the Hofstadter model~\cite{thouless1982,moller2009,moller2015}. Similar bands with high Chern number can also be generated in the TFB lattice model. Initially, $C=2$ TFB were constructed in a dice lattice model~\cite{wang2011a} and three-band triangular lattice model~\cite{wang2012}, and the systematic approaches to generate TFBs with arbitrary Chern number were further proposed in the multilayer systems~\cite{trescher2012} and the multi-orbital structures~\cite{yang2012a}. Although a $C>1$ Chern band can be mapped to a $C$-component Landau level using the hybrid Wannier states~\cite{barkeshli2012}, these components are mutually entangled with each other, in contrast with the case in the usual Halperin state. A series of fermionic fractional Chern insulators at $\nu=1/(2C+1)$ filling were established numerically in the TFB models with high Chern number $C$~\cite{wang2012,liu2012a,grushin2012,wang2013}, which can be understood as the $SU(C)$ color-entangled version of Halperin states~\cite{wu2013,wu2014a}. On the other hand, the topological states in bosonic system are much subtle. The bosonic fractional Chern insulators at $\nu=1/(C+1)$ filling were also proposed in the TFB models~\cite{wang2012,liu2012a,sterdyniak2013}. These incompressible states can also be understood by the concept of flux attachment in lattice system~\cite{kol1993,moller2009}, yielding a series of fractional Chern insultors at filling factors $\nu=r/(r\vert C\vert+1)$ for bosons, and $\nu=r/(2r\vert C\vert+1)$ for fermions with $r$ an integer~\cite{moller2015}. The bosonic integer quantum Hall (BIQH) state, a topological phase protected by $U(1)$ symmetry but in the absence of intrinsic topological order, was first perceived in the interacting two-component boson gases at the integer filling $\nu=1+1$~\cite{senthil2013,furukawa2013,wu2013a,regnault2013,geraedts2017}.
Its existence was also predicted by the composite fermion theory at integer filling $\nu=1$ of $\vert C\vert=2$ band ($r=-1$)~\cite{moller2015}, and numerically confirmed in different lattice models~\cite{sterdyniak2015a,he2015a,zeng2016,he2017a,andrews2018}.

So far, the topological nature of the Chern insulators follow the similar way established in the continuum limit with the Landau levels, i.e., the integer and fractional quantum Hall effect emerge at the integer and fractional filling. It is interesting to explore the novel fractional Chern insulators with no analogue in the continuum limit~\cite{moller2009,spanton2018}. In this paper, combining the exact diagonalization (ED) and infinite density matrix renormalization group (iDMRG) algorithm, we study the exotic fractional Chern insulators of the hard-core bosons in a specific multi-band model. The adopted triangular lattice model hosts two lower TFBs with respective Chern number $C=2$, and $-1$. We numerically confirm the previously proposed $\nu=1/3$ bosonic fractional Chern insulating (FCI) state. Its topological order is well characterized by the $\mathbf{K}=\big(\begin{smallmatrix}2&1\\1&2\\\end{smallmatrix}\big)$-matrix, i.e., a color-entangled $\frac{1}{3}$-FCI state with the two components originated from the high Chern number $C=2$ of the lowest TFB. More importantly, an unexpected bosonic FCI state, rather than the BIQH state, is observed at the integer fillings $\nu=1$ and $\nu=2$. The twofold ground-state degeneracy and the exact fractional-$1/2$ charge pumping, as well as the level counting of the low-lying entanglement spectrum, evidence the emergence of $\nu=1/2$ Laughlin-like FCI state. We identify that the two lower TFBs produce an effective $C=1$ Chern band with half-filling. In addition, we find a strict particle-hole-like symmetry between the $\nu$ and $3-\nu$ filling inherited from the intrinsic time-reversal symmetry of the specific model.

\begin{figure}[!htb]
    \begin{center}
        \includegraphics[width=0.8\linewidth]{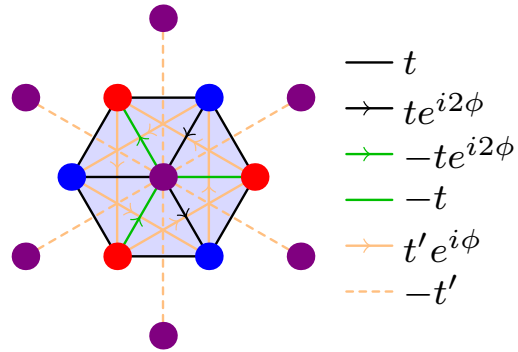}
    \end{center}
    \caption{(Color online). Schematic structure of the adopted triangular lattice model. The unit cell is highlighted by a gray hexagon, and the hopping process between the NN and NNN neighbors with specified gauge are given on the right hand.}
    \label{fig:lattice}
\end{figure}

{\it Model and method.---} We consider loading the hard-core bosons with $U(1)$ charge conservation into a specific three-band triangular-lattice model first proposed by one of our authors~\cite{wang2012}:
\begin{eqnarray}
H&=&\pm t\sum_{\langle\ij\rangle}
\left[b^{\dagger}_{i}b_{j}\exp\left(i\phi_{ij}\right)+\textrm{H.c.}\right]\nonumber\\
&&\pm t^{\prime}\sum_{\langle\langle ij\rangle\rangle}
\left[b^{\dagger}_{i}b_{j}\exp\left(i\phi_{ij}\right)+\textrm{H.c.}\right],
\label{e.1}
\end{eqnarray}
where $b^{\dagger}_{i}$ creates a hard-core boson at site $i$, $\langle\dots\rangle$ and
$\langle\langle\dots\rangle\rangle$ denote the nearest-neighbor (NN)
and the next-nearest-neighbor (NNN) pairs of sites with the specified phase factor denoted by the arrows (Fig.~\ref{fig:lattice}), respectively. Each unit cell (gray hexagon) contains three inequivalent sites, and thus three single-particle bands are created. The flatness ratio and topological index of each band are parameter-dependent (more details see Appendix~\ref{SP1}). Here, we fix $t=1$, $t^{\prime}=1/4$ and $\phi=\pi/3$~\cite{wang2012} unless specified, yielding $C=2$, $-1$, $-1$ in the each band (from bottom to top) and two lower flat bands with the respective flatness ratio about $15$, $14$, respectively. Those TFBs mimic the Landau levels in the continuum but with diverse Chern numbers. We, therefore, construct a multi-band model with high Chern numbers.

Numerically, we employ the unbiased ED and iDMRG methods to study the topological order of the potentially topological states. In ED scheme, a finite system of $N_{\text{orb}}=N_x\times N_y$ unit cells (the total number of site $N_{s}=3\times N_{x}\times N_{y}$) with periodic boundary condition in both directions is considered. The filling factor is defined as $\nu=N_{b}/N_{\text{orb}}$ with $N_{b}$ the number of hard-core bosons loaded. Due to the translational symmetry, the many-body eigenstates can be labeled by the total momentum quantum number $(k_x,k_y)$ in unit of $(2\pi/N_x, 2\pi/N_y)$. To access larger system sizes and extend numerical evidences beyond the ED method, we also study the interacting system on a infinite cylinder (with finite width $L_y$) by iDMRG alogrithm~\cite{white1992,mcculloch2008}. This method allows us to directly simulate the charge pumping process~\cite{grushin2015,he2014b,zaletel2014,zhu2015b} and conveniently explores the underlying topological order by the entanglement properties in the many-body ground state~\cite{li2008,luo2020}.

\begin{figure}[!htb]
    \begin{center}
        \includegraphics[width=0.45\linewidth]{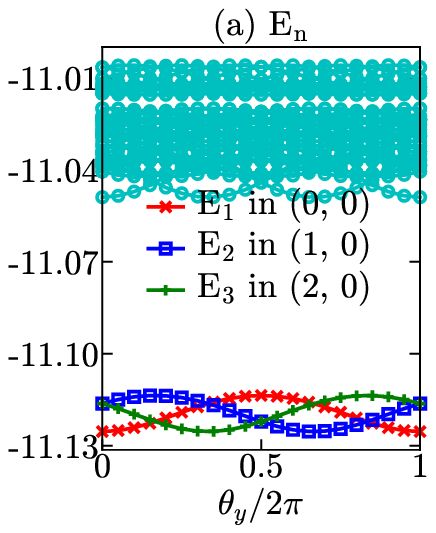}%
        \includegraphics[width=0.55\linewidth]{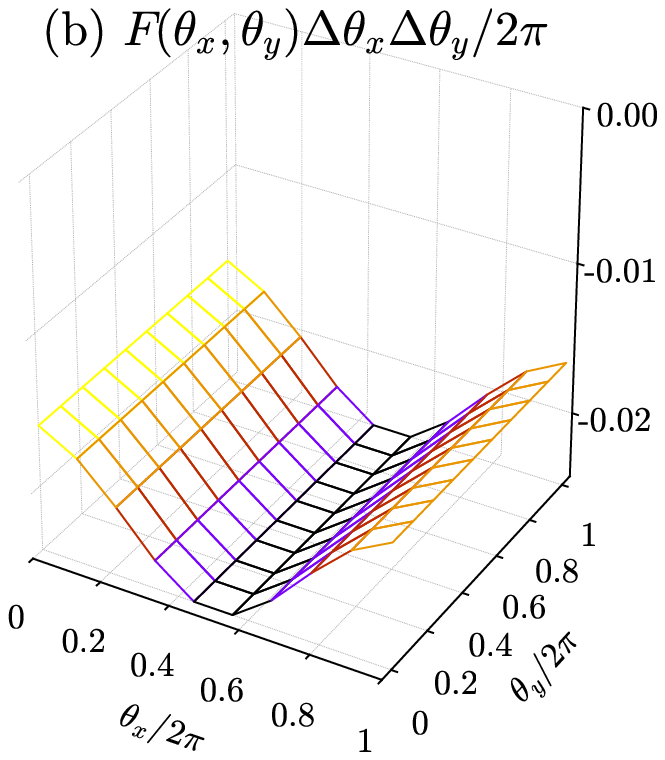}
        \includegraphics[width=\linewidth]{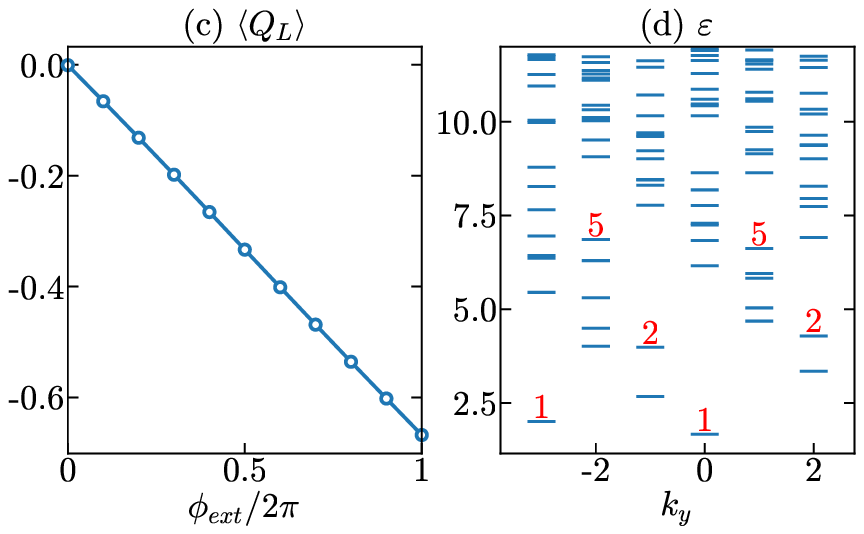}
    \end{center}
    \caption{(Color online). $\nu=1/3$ FCI state. (a) Spectral flow as a function of twist boundary phase $\theta_y$; (b) Total Berry curvature of the three quasi-degenerate ground states at $10\times10$ mesh points, which shares a total Chern number $C=2$; (c) $2/3$ charge pumping during the adiabatic flux quanta insertion; and (d) Momentum-resolved entanglement spectrum revealing two branches of chiral edge modes, each with degeneracy pattern: 1, 2, 5. The lattice size $N_{s}=3\times 3\times5=45$ is used in ED calculation. The cylinder width $L_{y}=6$ is adopted in iDMRG calculation.}
    \label{fig:BFQH_nu13}
\end{figure}

{\it FCI state at $\nu=1/3$.---}The $\nu=1/3$ bosonic fractional Chern insulator had been previously studied by the ED method on a torus geometry~\cite{wang2012,jaworowski2019}. The low-lying energy spectrum in Fig.~\ref{fig:BFQH_nu13} (a) unambiguously shows that three quasi-degenerate ground states, being robust against the boundary conditions, evolve into each other while are well separated from the higher excited states. By imposing the twisted boundary phases $\theta_x$ and $\theta_y$ in both directions, the Chern number of a many-body ground-state is given by $C =\frac{1}{2\pi} \int_0^{2\pi}\int_0^{2\pi} d\theta_x d\theta_y F(\theta_x, \theta_y)$, where $F(\theta_x, \theta_y) = \Im\left(\left\langle\partial_{\theta_x} \psi\middle\vert\partial_{\theta_y} \psi\right\rangle - \left\langle\partial_{\theta_y} \psi\middle\vert\partial_{\theta_x} \psi\right\rangle\right)$ is the gauge invariant Berry curvature with $\psi$ the ground-state wave functions~\cite{sheng2003}. The Berry curvature of the ground-state manifold [Fig.~\ref{fig:BFQH_nu13} (b)] contributes a total Chern number $C=2$. The topological order of $\nu=1/3$ FCI state can be well characterized by the $\mathbf{K}$ matrix in terms of Chern-Simons field theory~\cite{blok1990,wen1992b}. Here, the ground state degeneracy $d=\det \mathbf{K} = 3$ and the total Chern number $C=\sum_{i,j}(\mathbf{K}^{-1})_{ij}=2$, yielding $\mathbf{K}=\big(\begin{smallmatrix}2&1\\1&2\\\end{smallmatrix}\big)$.
Alternatively, this $\nu=1/3$ state is also predicted by composite fermion theory~\cite{moller2015} at bosonic filling $\nu=r/(r|C|+1)$ with $r=1$, where one flux quantum is attached to each boson to form the weakly interacting composite fermion. Incompressible state is expected to form when they fully fill integer number of Chern bands.

To complement the ED results on a torus, we study $\nu=1/3$ case on a cylinder geometry using the iDMRG algorithm.
The exact $2/3$ charge pumping after a flux quanta threading [Fig.~\ref{fig:BFQH_nu13}(c)] agrees with the Chern number calculation using ED method~\cite{wang2012}, where each ground state averages contributes to a Chern number $C=2/3$. Furthermore, two parallel propagating edge modes with the same level counting $(1, 2, 5, \cdots)$ are observed in the momentum resolved entanglement spectrum of the ground state [Fig.~\ref{fig:BFQH_nu13}(d)], in agreement with two positive eigenvalues in the $\mathbf{K}$-matrix description and the low-lying structure of edge excitations in $1/3$ FQH systems~\cite{li2008,zhu2016b}. Therefore, the present state is a two-component fractional quantum Hall state, in which the two-component nature originates from the $C=2$ Chern number of the lowest TFB. The exact equivalence between the two components render the bosoic FCI state a color-entangled $\frac{1}{3}$-FCI state~\cite{wu2013,wu2014a,jaworowski2019}, which differs from the general two-component $\nu=1/3+1/3$ case discussed previously in the Bose gas with two spin states~\cite{furukawa2012,wu2013a}, though they share the similar topological order.

{\it FCI state at $\nu=1$.---}It is commonly accepted that the integer quantum Hall effect emerges when electrons fully fill the Landau levels. The BIQH, characterized by $\mathbf{K}=\big(\begin{smallmatrix}0&1\\1&0\\\end{smallmatrix}\big)$, were predicted in the lowest Landau level filled with two-component interacting bosons at total filling $\nu=1+1$~\cite{senthil2013,furukawa2013}. This state was also proposed in the Hofstadter model when the $C=2$ Chern band are fully occupied by the hard-core bosons~\cite{zeng2016}. It is thus interesting to explore the potential state in the current lattice model at integer filling.

\begin{figure*}[!htb]
    \begin{center}
        \hspace{-0.2in}
        \includegraphics[width=0.18\linewidth]{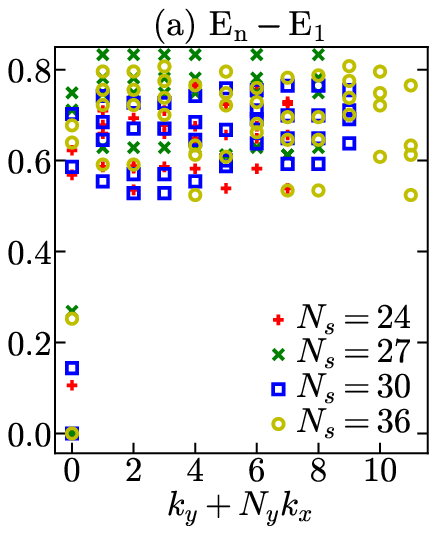} \ \
        \includegraphics[width=0.18\linewidth]{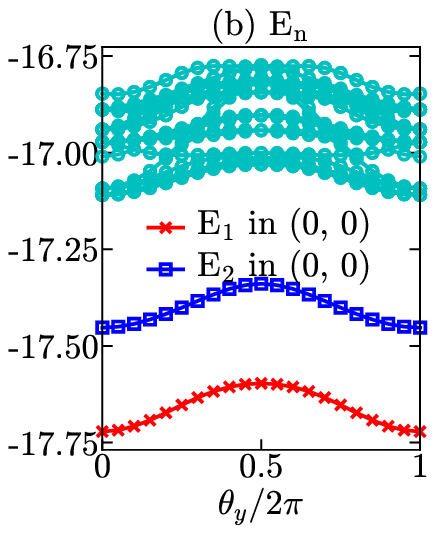}%
        \includegraphics[width=0.22\linewidth]{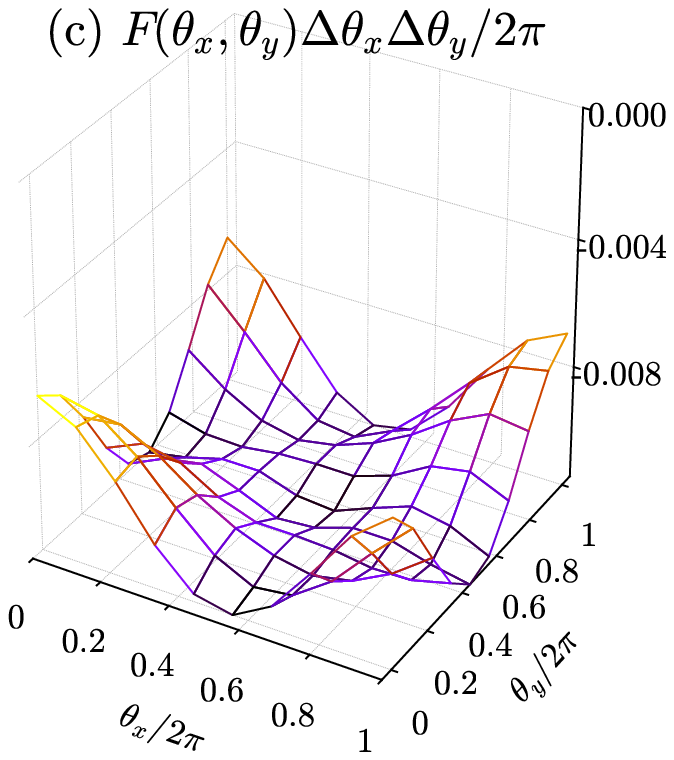}%
        \includegraphics[width=0.36\linewidth]{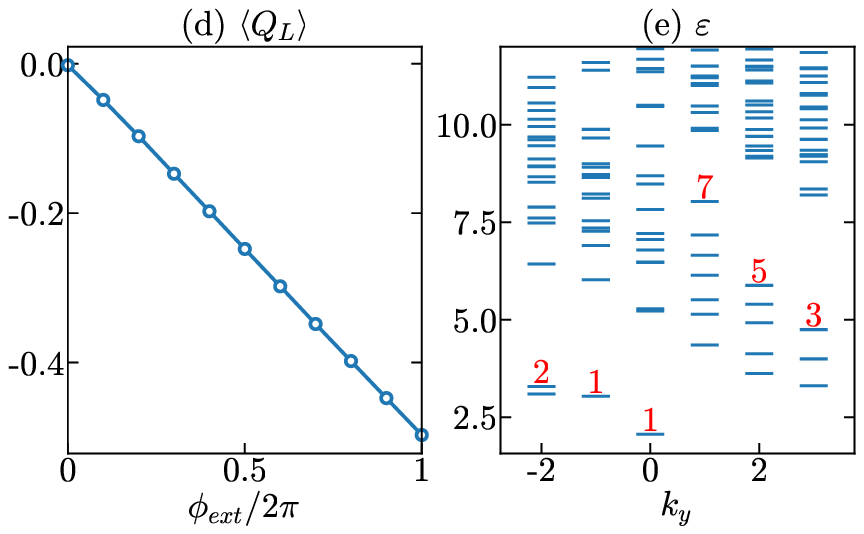}
    \end{center}
    \caption{(Color online). $\nu=1$ FCI state. (a) Low-lying energy spectrum for different lattice sizes; (b) Spectral flow as a function of twist boundary phase $\theta_y$; (c) Total Berry curvature of the two ground states, which share a total Chern number $|C|=1$; (d) $1/2$ charge pumping during adiabatic flux quanta insertion; (e) Momentum-resolved entanglement spectrum revealing one branch of chiral edge mode, with degeneracy pattern: 1, 1, 2, 3, 5, 7. The lattice size $N_{s}=3\times 3\times3=27$ is used in ED calculation. The cylinder width $L_{y}=6$ is adopted in iDMRG calculation.}
    \label{fig:BIQH}
\end{figure*}

We study the low-lying energy spectrum of different lattice sizes up to $36$ sites with the ED scheme, which is almost the current computational limit (Hilbert space dimension reaches $10^8$ for a single momentum sector). The twofold quasi-degenerate ground states, separated from the excited states by a finite gap, emerge at the fixed momentum sector $(k_x,k_y)=(0,0)$, irrespective of the lattice size [Fig.~\ref{fig:BIQH} (a)]. The robustness of the ground states manifold is also confirmed by the twisted boundary conditions $\theta_{x/y}$ [Fig.~\ref{fig:BIQH}(b)], where the two states never mix with higher excited states. This agrees with the previous statistical rule in $C=1$ TFB system~\cite{wang2011,wang2012,wang2012a,regnault2011}, where two ground states are predicted to emerge at $(k_x, k_y)$ and $(k_x+N_b, k_y+N_b) \mod (N_x, N_y)$ sector, respectively. In contrast, only a single ground state is observed in the $C=2$ Hofstadter bands at $\nu=1$~\cite{zeng2016}, in which a BIQH state is predicted. The FCI at integer filling $\nu=1$ is further supported by the iDMRG results. The charge pumping after a flux quanta threading is exactly $1/2$ [Fig.~\ref{fig:BIQH}(d)]. This agrees with the existence of two degenerate ground states carrying a total Chern number $C=1$. Moreover, only single branch of the edge mode is observed in the momentum resolved entanglement spectrum [Fig.~\ref{fig:BIQH}(e)], in sharp contrast with the usual situation of two branch of edge modes in a $C=2$ band. The corresponding counting sequence $(1, 1, 2, 3, 5, 7, \cdots)$ is consistent with that in a $1/2$ Laughlin state. Therefore, we find a $1/2$ Laughlin-like FCI state at the integer filling $\nu=1$, rather than the expected BIQH state.

The confirmed evidence of $1/2$ Laughlin-like fractional Chern insulator comes from the many-body Chern number of the ground-state wave function by integrating the Berry curvature in the first Brillouin zone. The discrete Berry curvatures~\cite{sheng2003} are relatively smooth [Fig.~\ref{fig:BIQH}(c)]. Their summations yield a precisely quantized Chern number $C_{tot}=1$, or averagely a $C=1/2$ fractional Chern number for each ground state. This is in stark contrast to the FCI states reported at the bosonic $\nu=1/3$ filling and the fermionic $\nu=1/5$ filling in this model before~\cite{wang2012} and the BIQH state reported in the bosonic gases~\cite{senthil2013,furukawa2013} and Hofstadter band~\cite{zeng2016}, where the ground state manifold contributes to a total Chern number $C_{tot}=2$. Considering the fact of two lower TFBs with $C=2$ and $C=-1$ in our model, it is natural to believe that the hard-core bosons occupy the two lower TFBs simultaneously, which effectively generate a $C=1$ topological band but with $\nu=1/2$ filling. It was previously reported the giving width to the Chern band with high Chern number can help to stabilize a fractional Chern insulator~\cite{grushin2012}. Here, the two TFBs, instead of high Chern number, play the similar role, which may be the reason why the $1/2$ FCI state remains robust at the integer filling though the effective flatness ratio of the generated $\vert C\vert=1$ Chern band is much reduced down to about $4$ (details see Appendix~\ref{SP1}). The less strict condition for the fractional Chern insulator may provide new platform to realizing the fractional quantum hall state in multi-band systems.

\begin{figure}[!htb]
    \begin{center}
        \includegraphics[width=\linewidth]{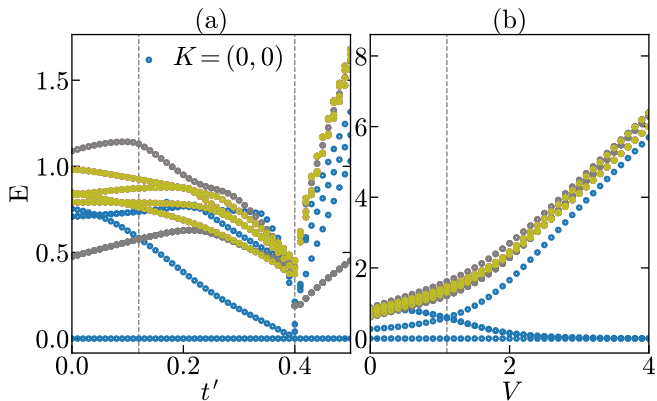}
    \end{center}
    \caption{(Color online). Robustness of $\nu=1$ FCI state. Evolution of low-lying energy spectrum versus NNN hopping strength $t^\prime$ (a) and NN interaction $V$ (b). System size is $N_s=3\times3\times3=27$. Colors label distinct momentum sectors, in particular states at $(0,0)$ momentum sector are colored blue. In (a), the $\nu=1$ FCI state is dominated between $0.12<t^\prime<0.4$, where the two ground states always share a Chern number $|C|=1$. Beyond this region, two topologically trivial states  emerge. In (b) a phase transition from the $1/2$ Laughlin-like FCI state to the topologically trivial charge-density-wave state occurs at $V=1.1$ at integer fillings $\nu=1$.}
    \label{fig:evolution}
\end{figure}

{\it Robustness of FCI state at $\nu=1$.---}Similar band topology but in the absence of the flat bands were previously constructed in a specific Hofstadter lattice model~\cite{wang2013}, where the BIQH state is predicted at integer filling with the hard-core bosons~\cite{zeng2016}. It is therefore interesting to study whether the FCI state at integer filling turns into the BIQH state, or other competing order, when the flatness of TFBs is lowered. The flatness of TFBs is closely related to the NNN hopping integral $t^{\prime}$ (Appendix~\ref{SP1}). We show the low-energy many-body spectrum $E_{n}-E_{1}$ evolving with $t^{\prime}$ in Fig.~\ref{fig:evolution} (a) on a torus with fixed $N_{s}=3\times3\times3=27$ sites. The twofold ground state quasi-degeneracy remains evident among $0.12\leq t^{\prime}\leq 0.4$, indicating the robustness of FCI state at $\nu=1$ filling in this region. When $t^{\prime}$ is reduced down to below $0.12$, only a single ground state separated by excited states is observed. In contrast to BIQH, this is a trivial state with $C=0$. Another phase transition is observed at $t^{\prime}=0.4$, where the underlying band topology changes (Appendix~\ref{SP1}). We do not find any signature of the BIQH state within the current setup.

On the other hand, the FCI state is often challenged by other conventional ordered states in the strongly correlated systems~\cite{sheng2011}, e.g., the charge-density wave state may be favorable when the long-range Coulomb interaction are taken into account~\cite{wang2011,luo2020}. We further discuss the role of NN Coulomb repulsion $V$~\cite{zeng2020a} on the FCI state. In the large $V$ limit, the loading bosons tend to occupy the NNN neighbors, i.e, the same sublattice, to lower energy contributed by the Coulomb repulsion. Fig.~\ref{fig:evolution} (b) shows a phase transition occurring at $V_{c}\simeq1.1$. The FCI state at the integer filling $\nu=1$ is robust below $V_{c}$ since the two twofold quasi-degeneracy remains. Above the critical $V_{c}$, a topologically trivial charge-density-wave state characterized by the threefold quasi-degenerate ground state emerges, in agreement with the constructed three sublattices in our model.

{\it Particle-hole symmetry.---}Since no band projection is performed here, we are able to study the system at higher fillings with $\nu>1$. Giving the unexpected FCI state at $\nu=1$ filling, in which an effective $\frac{1}{2}$-filled Chern band with $C=1$ is produced. Is it possible that the BIQH state emerges when the two lower TFBs are fully occupied? Our numerical results on $\nu=2$ filling show that the present bosonic system remains the $1/2$ Laughlin-like FCI state. We, therefore, establish a fractional Chern insulator at the integer fillings in our constructed bosonic lattice model. It should be emphasized that the energy spectrum at $\nu=2$ filling is exactly the same as that at $\nu=1$ filling. The charge pumping, and entanglement spectrum, except for a minus sign, and the reversal propagating direction, respectively, are also the same (Appendix~\ref{SP2}). Similar behavior can be further observed at the filling of $\nu=\frac{1}{3}$ and $\nu=\frac{8}{3}$. It seems that there exist an particle-hole-like symmetry between $\nu$ and $3-\nu$ filling in our adopted model. The particle-hole symmetry between the $\nu$ and $1-\nu$ filling of fermions and between $\nu$ and $2-\nu$ filling of two-component bosons in the quantum Hall state was previously reported when particles partially occupy the lowest Landau level~\cite{geraedts2017,pan2020}, where the particle-hole symmetry is preserved due to the perfectness of the Landau level. Here, no such symmetry explicitly exists as revealed by the band structure. Considering a particle-hole transformation $\mathcal{P}_{\text{ph}}$ i.e., $b^{\dag}_{i}\leftrightarrow b_{i}$, yielding $\mathcal{P}_{\text{ph}}H(\phi)\mathcal{P}^{-1}_{\text{ph}}=H(-\phi)$. We show that the only difference between the $\phi$ and $-\phi$ is the reversed sign of the Chern number for each bands. Therefore, an additional particle-hole-like symmetry between $\nu$ and $3-\nu$ filling exists in our constructed bosonic lattice model. Such symmetry originates from the intrinsic commute relation of bosons and the special time-reversal symmetry of the model Hamiltonian (more details see Appendix~\ref{SP2}). We emphasize that no particle-hole symmetry is presented in the corresponding fermionic system.

{\it Summary and discussion.---}In summary, we study the fractional Chern insulating states of the hard-core bosons on a specific three-band triangular lattice model possessing two lower TFBs with $C=2$ and $C=-1$. Combining the ED and iDMRG algorithm, we characterize the underlying topological orders by ground state degeneracy, low energy spectrum, Chern number, together with charge pumping and entanglement spectrum. At $\nu=1/3$ filling, a color-entangled $\frac{1}{3}$ FCI state is observed, in which two components come from the Chern number $C=2$ of the lowest TFB. Particularly, instead of a BIQH state, we find a robust $1/2$ Laughlin-like fractional Chern insulating state at the integer fillings. The loading hard-core bosons simultaneously occupy the two lower TFBs, producing an effective $C=1$ Chern band with $\nu=1/2$-filling. In addition, a strict particle-hole-like symmetry between the $\nu$ and $3-\nu$ fillings, originated from the intrinsic commute relation of bosons and the time-reversal symmetry of the specific model, is revealed in present lattice model. Our results demonstrate the rich family of the novel topological states in the lattice systems with high Chern number and multi-bands, and add the new insight into the FCIs.

Further understanding the topological order of the present FCI state at integer fillings, especially constructing a model fractional state and performing the overlap with the ground state wave function obtained by the ED method, is highly desirable. Exploring more exotic FCI states and potential relation to the generalized Pauli principle in the multi-band system with nontrivial topology should be interesting for future study. Such multi-band physics is neglected in most previous studies due to the simplified band projection. Whether the multi-band induced fractional Chern insulators can also be realized in the fermionic system is also interesting. Very recently, a direct transition between $\frac{1}{3}$ Laughlin state and Chern insulating state driven by the interaction is reported in the Harper-Hofstadter model loaded by fermions~\cite{schoonderwoerd2019}, where the lowest Landau level comprises multiple sub-bands. This may be the clue of multi-band and high Chern number induced fractional state at integer filling in fermionic system.

The single-occupancy constraint of hard-core bosons may prevent the emergence of BIQH state in present lattice model. Whether the BIQH state will survive in this TFB model loaded by soft-core bosons, especially for intermediate interaction regime, is also an interesting problem that worth delicate investigation. A first study of three-body hard-core bosons with intermediate Hubbard interaction strength has not given any clue of BIQH state. Since the computational effort to study the full phase diagram in current TFB model with soft-core bosons are much larger than that of hard-core bosons, and more exotic non-Abelian FCI phases that occupy multiple Chern bands may also emerge, we will leave the numerically intensive problem in a detailed future work.

{\it Acknowledgments.---} We thank S.-L. Yu, W. Zhu, and T.-S. Zeng for helpful discussions. This work is supported by the National Nature Science Foundation of China Grant Nos. 12074175 and 11874325. YZ and CDG also acknowledge the Ministry of Science and Technology of China under Grant No.
2016YFA0300401. ALH acknowledges the China Postdoctoral Science Foundation (Nos. 2020M680499).

\section*{Appendix For ``Bosonic fractional Chern insulating state at integer fillings in multi-band system''}
\setcounter{figure}{0}
\setcounter{equation}{0}
\renewcommand \thefigure{S\arabic{figure}}
\renewcommand \theequation{S\arabic{equation}}

The adopted triangular lattice model loaded with the hard-core bosons has been schematically shown in the main text. The model Hamiltonian can be written as~\cite{wang2012}
\begin{eqnarray}
H&=&\pm t\sum_{\langle\ij\rangle}
\left[b^{\dagger}_{i}b_{j}\exp\left(i\phi_{ij}\right)+\textrm{H.c.}\right]\nonumber\\
&&\pm t^{\prime}\sum_{\langle\langle ij\rangle\rangle}
\left[b^{\dagger}_{i}b_{j}\exp\left(i\phi_{ij}\right)+\textrm{H.c.}\right].
\label{es.1}
\end{eqnarray}
Each unit cell comprises three inequivalent sites, which naturally produce three single particle bands. Only the nearest-neighbor (NN) and the next-NN hopping process of the hard-core bosons are considered. It should be reminded that no on-site term exist in our model, which is crucial for the for the particle-hole-like symmetry existing in the present bosonic model as discussed below. In addition, we restrict our model in the canonical ensemble, which guarantee the particle number conservation. Even no explicit interactions are considered, the model system remains strongly correlated due to the nature of hard-core boson.

\section{Single-particle energy spectrum} \label{SP1}
\begin{figure}[!htb]
    \begin{center}
        \includegraphics[width=\linewidth]{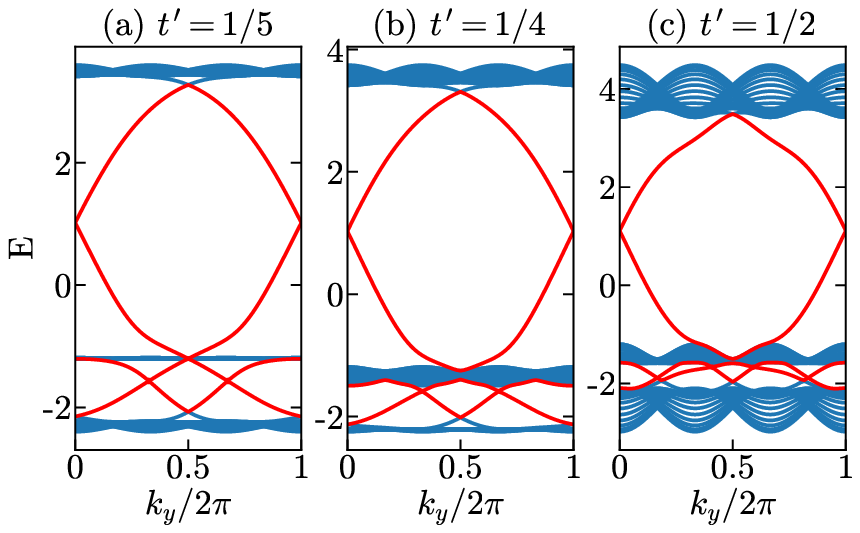}
    \end{center}
    \caption{(Color online). Single-particle spectrum of the three-band triangular lattice model for different $t^{\prime}$ at fixed $\phi=\pi/3$. The Chern numbers, from low to high energy bands, for $t^{\prime}<1/3$ topological phase is $(2,-1,-1)$. (c) Higher-Chern-number bands exists for $t^{\prime}>1/3$, which is $C=(-4,5,-1)$ at $t^{\prime}=1/2$.}
    \label{sfig:spectrum}
\end{figure}

The band topology of the present lattice model are parameters dependent. In Fig.~\ref{sfig:spectrum}, we show the band structure on cylinder evolving with the next-NN hopping integral $t^{\prime}$ at fixed $\phi=\pi/3$. A topological phase transition occurs at $t^{\prime}_{c}=1/3$. Below $t^{\prime}_{c}$, the Chern number for the respective Chern band is $(2, -1, -1)$ (from bottom to top), while it turns to be $(-4, 5, -1)$ for $t^{\prime}>t^{\prime}_{c}$. For large enough $t^{\prime}>1$, a topological phase with Chern number $(5,-4,-1)$ may also exist. We emphasize that there exist a special time-reversal-like symmetry in the present model. The band structure remains unchanged except for the reversed Chern number for each band when $\phi\rightarrow -\phi$ but with the other parameters fixed. The chiral edge states shown in Fig.~\ref{sfig:spectrum} reverse their propagating direction (not shown).

\begin{figure}[!htb]
    \begin{center}
        \includegraphics[width=\linewidth]{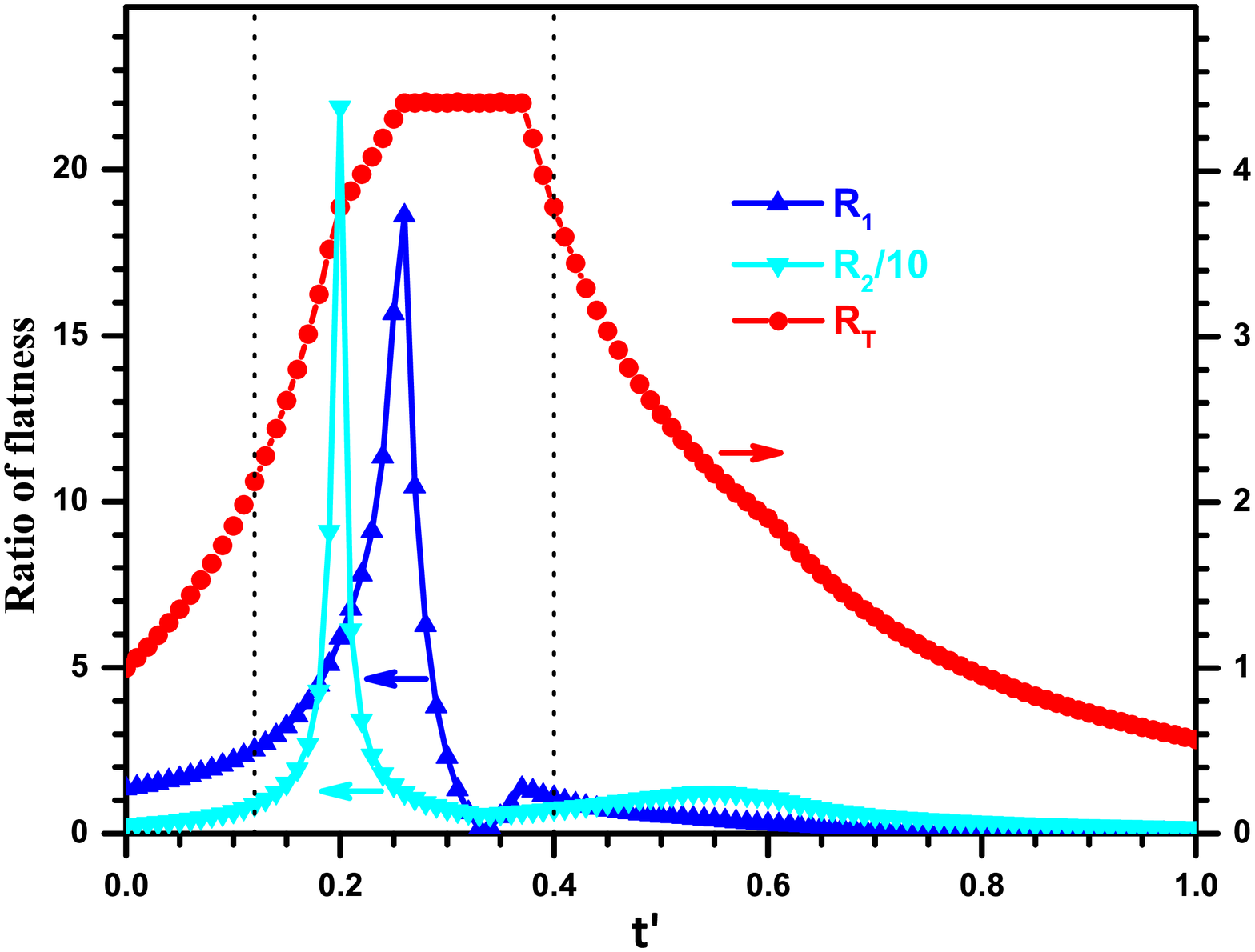}
    \end{center}
    \caption{(Color online). $t^{\prime}$-dependence of the flatness ratio of the respective topological flat band, together with that of the generated $C=1$ Chern band. The flatness ratio of the middle topological flat band is renormalized by a factor of $10$. $\phi$ is fixed at $\pi/3$.}
    \label{sfig:flatness}
\end{figure}

On the other hand, the flatness ratio of the respective topological band also varies with parameters as shown in Fig.~\ref{sfig:flatness}. We define the flatness ratio of the respective lower topological band by $\Delta_{i}/W_{i}$, where $\Delta_{i}$, and $W_{i}$ denotes the band gap above, and the bandwidth of, the selected $i$-th band ($i=1,2$ from bottom to top), respectively. With fixed $\phi=\pi/3$, the flatness ratio of the lowest topological flat band (TFB) reaches its maximum at $t^{\prime}=\frac{1}{4}$. It decreases monotonically with enhanced or weakened $t^{\prime}$. The flatness ratio of the middle TBF could be higher than $200$ at $t^{\prime}=\frac{1}{5}$. As mentioned in main text, the two lower TFBs jointly generate an effective $C=1$ Chern bands at integer fillings, which is the reason of the emergence of $\frac{1}{2}$ Laughlin-like FCI state at integer filling. We also show the flatness ratio of the generated effective $C=1$ Chern band comprising two lower TFBs, defined as the $R_{T}=\Delta_{2}/(W_{1}+W_{2}+\Delta_{1})$. The flatness ratio of this Chern band in the region where the $\frac{1}{2}$ Laughlin-like FCI state at integer fillings is observed is relative high, but much reduce in comparison with that of the respective TFB. In this sense, the multi-band physics provides the easy access to the fractional Chern insulating state.

\begin{figure}[!htb]
    \begin{center}
        \includegraphics[width=\linewidth]{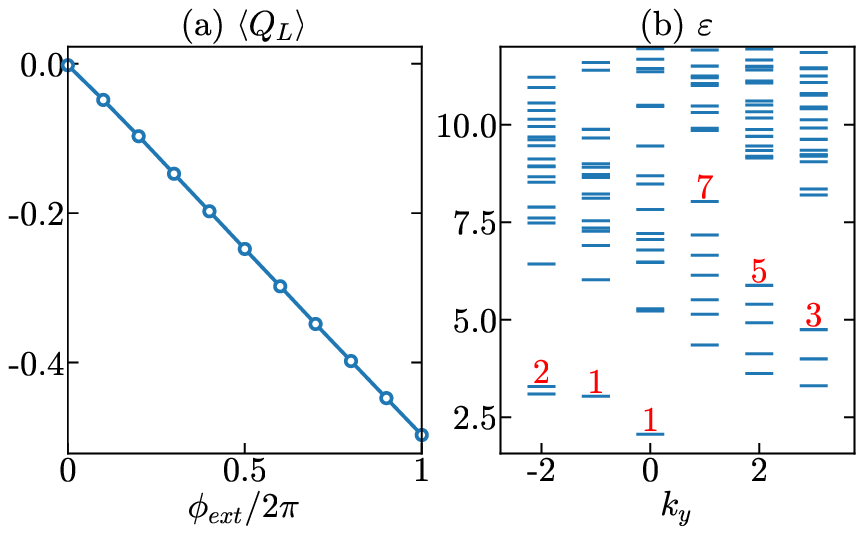}
        \includegraphics[width=\linewidth]{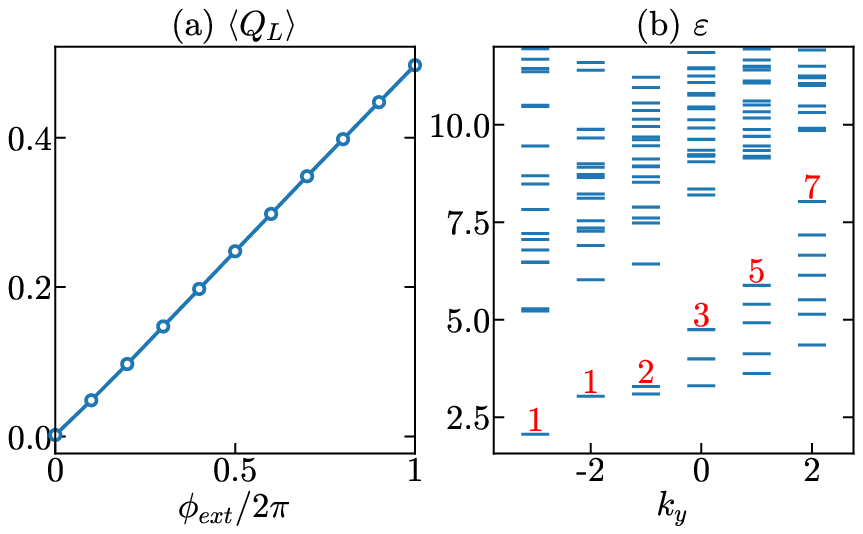}
    \end{center}
    \caption{(Color online). $\nu=1$ (Top) and $\nu=2$ (bottom) FCI states. (a) $1/2$ charge pumping after one flux insertion. (b) Momentum-resolved entanglement spectrum reveals one branch of chiral edge mode with degeneracy pattern: 1, 1, 2, 3, 5, 7, ... The two FCI states are time-reversal counterparts of each other, where the charge pumping differs by a sign and edge mode differs by the propogating direction.}
    \label{Sfig:BIQH_symmetry}
\end{figure}

\section{Particle-hole-like symmetry} \label{SP2}
Obviously, the band structure in the above section shows no particle-hole symmetry for fermions. However, the situation in bosonic system is significantly different from that in the fermionic system due to the intrinsic statistical law. In a hard-core bosonic system, we apply the particle-hole transformation $\mathcal{P}$, i.e., $b^{\dag}\leftrightarrow b$, on Hamiltonian (\ref{es.1}), yielding
\begin{eqnarray}
H&=&\pm t\sum_{\langle\ij\rangle}
\left[b_{j}b^{\dagger}_{i}\exp\left(i\phi_{ij}\right)+\textrm{H.c.}\right]\nonumber\\
&&\pm t^{\prime}\sum_{\langle\langle ij\rangle\rangle}
\left[b_{j}b^{\dagger}_{i}\exp\left(i\phi_{ij}\right)+\textrm{H.c.}\right].
\label{es.2}
\end{eqnarray}
Due to the commute relation between $\left[b_{i},b_{j}^{\dag}\right]=0$, we have $\mathcal{P}H(\phi)\mathcal{P}^{-1}=H(-\phi)$. We have shown that the band structure remains unchanged except for the reversed Chern number for $\phi\rightarrow -\phi$. Therefore, the only change for topological state is the reversed propagating direction, or \emph{particle} to \emph{hole}. Such particle-hole symmetry may be broken when the mass terms, i.e., $b^{\dag}_{i}b_{i}$, are taken into account, by which $\left[b_{i},b_{i}^{\dag}\right]=1$. Therefore, the particle-hole-like symmetry found in the present bosonic lattice model originates from the intrinsic commute relation of the bosons and the unique time-reversal symmetry of the specific model. This symmetry is naturally absent in the fermionic system due to the anti-commute relation of fermionic operators.

The particle-hole-like symmetry may be further understood by mapping the hard-core bosons onto the spin-$\frac{1}{2}$ quantum model. Using the Matsuda-Matsubara transformation, $b^{\dag}_{i}\rightarrow S_{i}^{+}$, $b_{i}\rightarrow S_{i}^{-}$ and $n_{i}\rightarrow S_{i}^{z}+\frac{1}{2}$, the above Hamiltonian can be expressed as
\begin{eqnarray}
H&=&\pm t\sum_{\langle\ij\rangle}
\left[S^{+}_{i}S^{-}_{j}\exp\left(i\phi_{ij}\right)+\textrm{H.c.}\right]\nonumber\\
&&\pm t^{\prime}\sum_{\langle\langle ij\rangle\rangle}
\left[S^{+}_{i}S^{-}_{j}\exp\left(i\phi_{ij}\right)+\textrm{H.c.}\right].
\label{es.3}
\end{eqnarray}
The particle-hole transformation in the bosonic model is same as the spin flip process in spin model. In the absence of symmetry breaking term $S_{i}^{z}$, the above Hamiltonian after spin flip also results in $H(-\phi)$.

In Fig.~\ref{Sfig:BIQH_symmetry}, we plot the charge pumping and entanglement spectrum to show the particle-hole-like symmetry between the $\nu=1$ and $\nu=2$, in which a $1/2$ Laughlin-like fractional Chern insulating state is realized at integer fillings. The two cases are almost exactly same, except for the charge (or hole) pumping process and reversal propagating direction. Similar behavior can be also found at the integer filling $\nu=1/3$ and $\nu=8/3$, where a color-entangled $\frac{1}{3}$ FCI state is observed.

\bibliography{FQHE_C2}

\end{document}